\documentclass[11pt]{amsart}
\usepackage{color}
\usepackage{graphicx}
\usepackage{amssymb}
\usepackage{amsbsy}
\usepackage{caption}
\usepackage[total={6in,9in},
top=1in, left=1.3in, right=1in, bottom=1in]{geometry}

\title{The Polyacetylene Raman Spectrum, Decoded}
\author{Eric J. Heller$^{1,2}$, Yuan Yang$^2$, and Lucas Kocia$^2$}
 
\begin{document}
\maketitle
\captionsetup{width=.9\textwidth}
 
$^1$Department of Physics, Harvard University, Cambridge, MA 02138

$^2$Department of Chemistry and Chemical Biology, Harvard University, Cambridge, MA 02138
 
\maketitle

 \vskip .2in
{\bf More than 30 years ago, polyacetylene was very much in the limelight,   an early example of a conducting polymer and  source of many unusual  spectroscopic features spawning    disparate ideas as to their origin.   Several versions of the polyacetylene spectrum story emerged,  with      contradictory conclusions.  In this paper both ordinary and peculiar polyacetylene spectral features  are explained in terms of standard (if disused) spectroscopic concepts, including the dependence of electronic   transition  moments    on phonon coordinates, Born-Oppenheimer   energy surface properties,  and (much more familiarly) electron and  phonon band structure.    Raman sideband dispersion  and line shapes   are very well matched by  theory in a fundamental way.    Most importantly,  clear ramifications  emerge for the Raman spectroscopy of a wide range of    extended systems, including graphene and beyond, suggesting changes to some common practice in condensed matter spectroscopy.    }

 \vskip .2in
 
\begin{figure}[htbp] 
   \centering
   \includegraphics[width=5.5in]{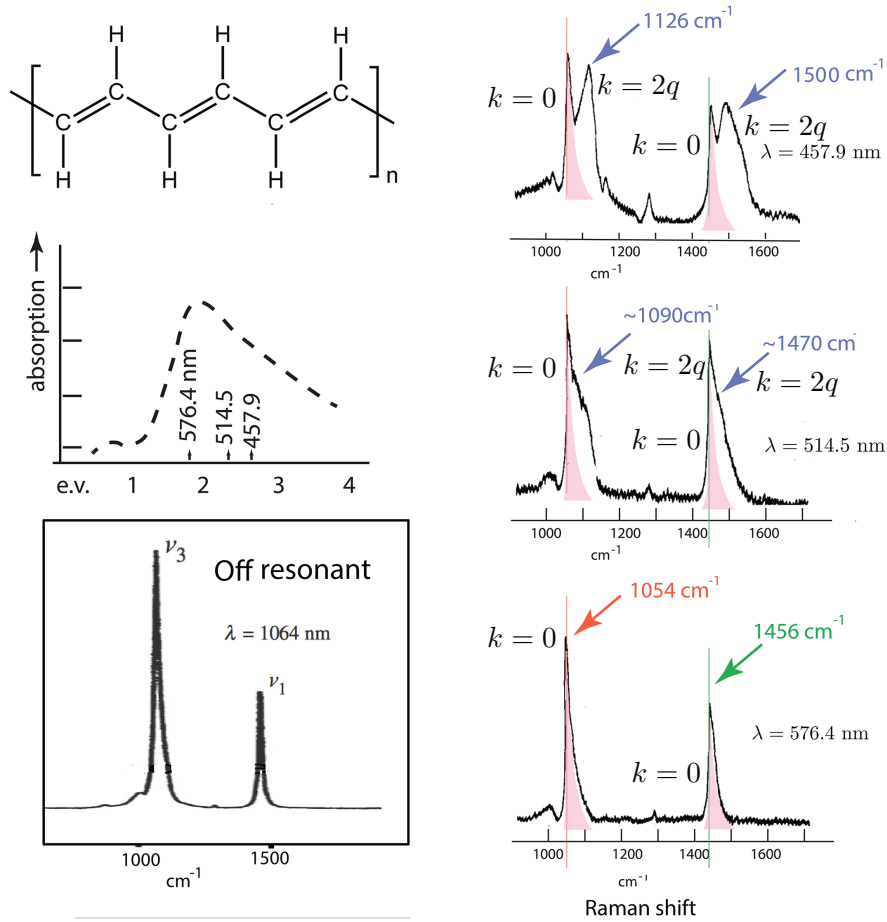}
   \caption{Some of the unusual spectroscopy of polyacetylene. (Top left) The structure of polyacetylene, with alternating single and double bonds, the result of a Peirls instability.  The double bonds are shorter 
   although the $\pi$ electrons are delocalized over the whole molecule. (Middle left) Absorption spectrum of polyacetylene in the region investigated for resonance Raman scattering.  (Bottom left): Off-resonance Raman spectrum of polyacetylene from reference \cite{zerbi}. (Right column) Some of the major features and changes that need an explanation, as incident wavelength decreases, here from 576.4, then 524.5, to 457.8 nm. 
   Red and green arrows point to sharp, nearly one-sided bands (shaded pink, tailing off to the right) that  show no dispersion or growth with incident $\lambda$. Blue arrows point to dispersive bands growing  in strength and also developing increasing frequency displacement from neighboring $k=0$ fixed  bands (dispersion) with increasing photon energy, still quite peaked. Spectral data on the left and middle right are re-drawn from reference~\cite{jumeau}; }
   \label{fig:poly}
\end{figure}

The polyacetylene molecule (figure~\ref{fig:poly}) once  played an outsized role, first   as a promising  organic conducting polymer\cite{Heeger}, then the focus of the Su-Schrieffer-Heeger\cite{SSH,kivelson,nat} model for soliton behavior of the Pierls distortion of the chain.  Around the same time, intensive work on its spectroscopy, especially Raman spectroscopy, was begun\cite{Kuzmany}.   Heeger, MacDiarmid, and Shirakawa shared the Nobel Prize in Chemistry in 2000 ``for the discovery and development of conductive polymers", notably polyacetylene. 
The polyacetylene spectroscopy boom   trailed off rather inconclusively.  Unusual  spectral features were assigned anticlimactically  to polydisperse   samples\cite{junk1}, unconventional vibrational patterns\cite{junk2}, co-existence of  ordered and a disordered phases  (another kind of polydisperse sample)\cite{junk3}.  Solitons\cite{junk2,junk3}, important as they were for other reasons, were cited as the cause  of the  signature    Raman scattering effects, an idea that never stuck.    Here we show the spectral features,  across a wide range of  experiments,   in fact are attributable to the internal  quantum dynamics of monodisperse samples.

Today, graphene is the new polyacetylene,  so to speak\cite{genref,dressel,natmat}. Strong  scientific, programmatic, and historical analogies  exist between the two, including an enigmatic    Raman spectrum, hopes for new devices based on conducting organic crystals, and  Nobel prizes for making and understanding a new ``molecule'' with amazing properties\cite{natmat}. 

Polyacetylene  is   simpler but  still very analogous to graphene.  
It therefore struck us as odd that polyacetylene's Raman spectrum remained mysterious, while graphene   has enjoyed a well established narrative,   for the past 12 years\cite{genref,dressel}.     
  
\textbf{Making full use of Franck-Condon theory}  The  key  to decoding the polyacetylene spectrum relies on a little used aspect of solid state Franck-Condon theory:   the nuclear coordinate dependence of the transition moment $\mu({ \boldsymbol \xi})$, where ${ \boldsymbol \xi}$  represents phonon coordinates.      
   The transition moment $\mu({ \boldsymbol \xi})$  controls  the amplitude to absorb a photon as a function of nuclear geometry, defined by the phonon coordinates ${ \boldsymbol \xi}$,  and is an integral part of  Franck-Condon theory (see below).   Neglecting the  nuclear coordinate dependence (the Condon approximation)     is  common almost to the point of universal, but we believe is unjustified at least  in systems consisting of   large  networks of conjugated carbon.  This was recently emphasized  by Duque \textit{et. al.}, who found the coordinate dependence was needed to explain spectroscopic data  for carbon nanotubes\cite{condonfail}.    

 The second order perturbation theory Franck-Condon  approximation is exactly equivalent to Born-Oppenheimer theory plus light-matter perturbation theory\cite{leeheller1,heller2}. It carries  two intrinsic phonon generation mechanisms, namely  (1) nuclear forces changes leading to motion on the excited electron Born-Oppenheimer state  (this may not be a factor in large, conjugated systems; see below), and (2) instantaneous phonon production via the coordinate dependance of the transition moment: $\mu({ \boldsymbol \xi}) \phi_0({ \boldsymbol \xi}) = \left [\mu_0 + \sum\limits_{j,k} (\partial \mu/\partial \xi_{j,k})\cdot  \xi_{j,k}+ \cdots\right ] \phi_0({ \boldsymbol \xi}) = a \phi_0({ \boldsymbol \xi})   + \sum\limits_{j,k} b_{j,k} \phi_{\xi_{j,k}}({ \boldsymbol \xi}) + \cdots $, where $\phi_{\xi_{j,k}}({ \boldsymbol \xi}) $ is a one-phonon mode  of band $j$ and   Bloch vector $k$, etc.


    \textbf{No change in potential upon photo absorption}  Due to dilution of the delocalized $\pi$ orbital amplitude over the infinite chain,  it is almost obvious that the Born-Oppenheimer potential energy surface is  unchanged after  one electron-hole pair formation in a system with a huge number of delocalized orbitals. (Longer times, sometimes too long to matter to Raman scattering,   find spontaneous localization taking place\cite{shank}).   The stability of the Born-Oppenheimer potential  was discussed by Zade\cite{zade} \textit{ et. al.}  where the  reorganization energy (the potential energy available    in the excited state starting at the ground state geometry) 
    in  long linear oligothiophenes was shown to decrease to zero with polymer length N. This stability means phonons are to lowest order created only through the coordinate dependence of $\mu({ \boldsymbol \xi}) $.  Thus the potential does not change in the excited state and  should not be held responsible  for electron-phonon scattering.  The transition moments and their coordinate dependence instead explain the phonon production. 
    
    \textbf{Contrast with double resonance}    The popular ``double resonance''  approach  came from a very general formalism of Falicov\cite{mf}  that did not suppose even the Franck-Condon (or the Born-Oppenheimer) approximation  and thus resorted to further (beyond second order) perturbation theory to sort things out. However if   Franck-Condon, Born-Oppenheimer theory is used it is not at all clear a retreat to perturbation theory is still necessary in the Falicov sense, as has become so popular in the ``double resonance'' approach to graphene spectroscopy\cite{dr}.    We know far too much about the input to Franck-Condon theory to abandon it to perturbation theory.  As has been shown, the potentials don't change in the excited states, and transition moment coordinate dependence can be strong and will not be  a mystery to modern electronic structure codes.

\begin{figure}[htbp] 
\advance\leftskip-1cm
 \includegraphics[width=7in]{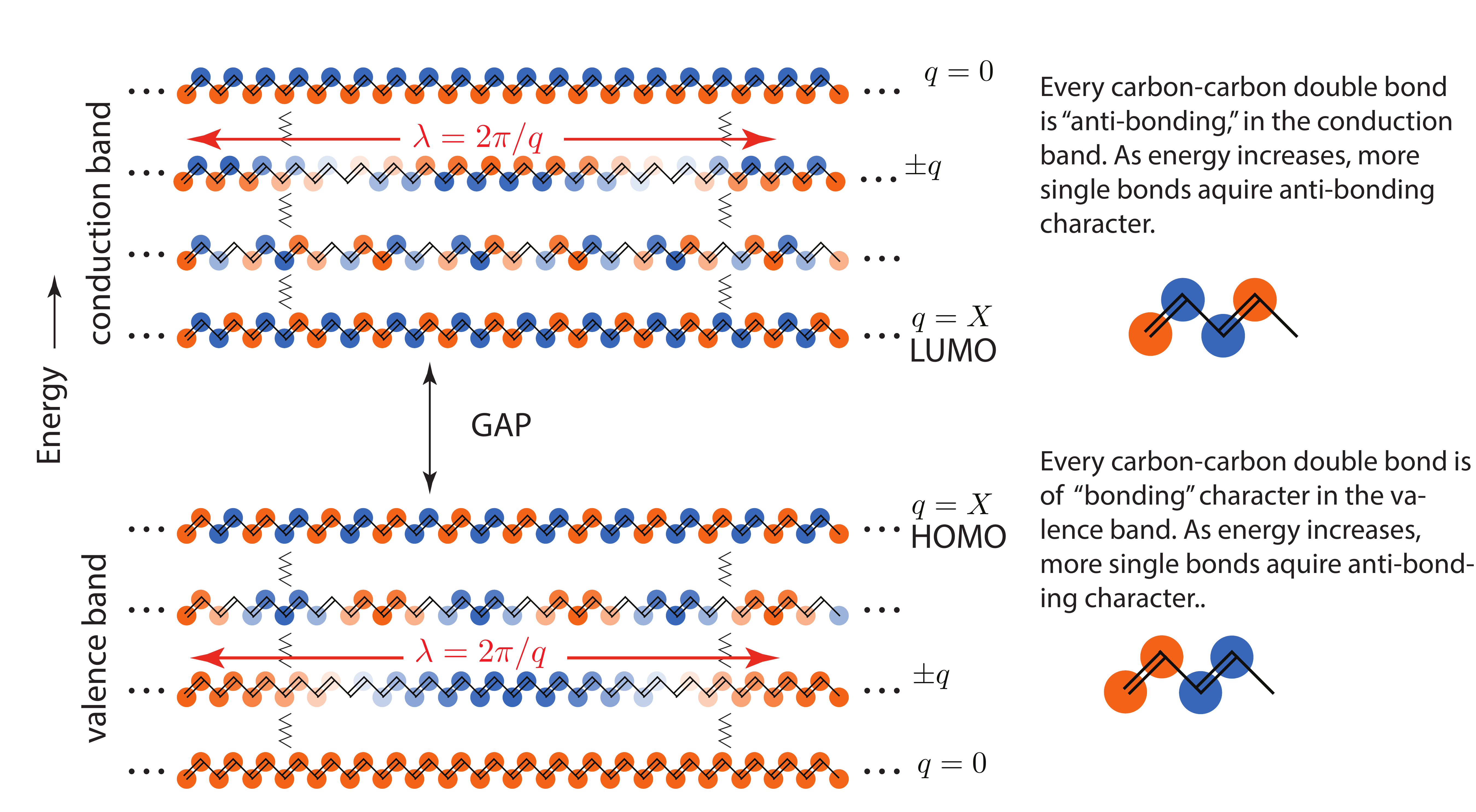} 
   \caption{Schematic of the  $\pi$ orbitals in an infinite polyacetylene chain. Each colored dot represents the top half of a carbon  $p_z$ orbital, with two colors giving the sign of that lobe of the orbital. The hidden lobe of each  orbital in this top down view is   below the plane of the molecule and of opposite sign to the visible part. Color intensity depicts wave function amplitude.  In the valence band, all double bonds are of bonding character (same sign on adjacent carbons).  In the conduction band, all double bonds are of anti-bonding character.
 }
   \label{fig:polyacetylene1}
\end{figure}

\textbf{Tight binding model} We present a schematic polyacetylene tight binding calculation based on out of plane carbon $p_z$ orbitals.  The model  makes the subsequent quantitative discussion justifying the qualitative theory much easier to understand and   keep short. 

\begin{figure}[htbp] 

 \advance\leftskip-1cm
   \includegraphics[width=7in]{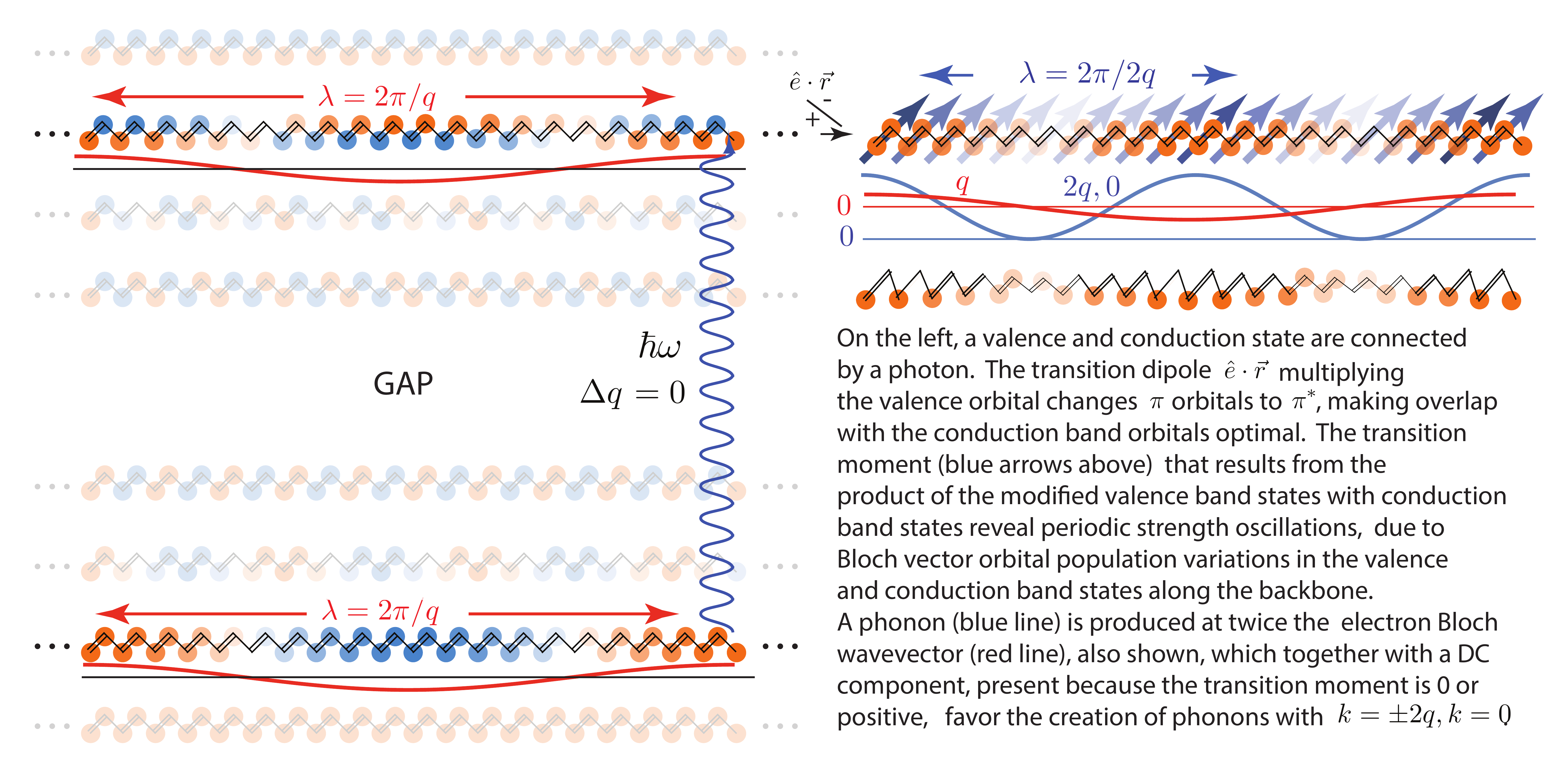} 
   \caption{A photon  creates an electron-hole pair, promoting an electron in a valence orbital to a conduction orbital  of the same $q$ to conserve crystal momentum, (or of opposite $q$, also  conserving crystal momentum, if the birth is accompanied by a creation of  a phonon of wave vector $2q$ via a Herzberg-Teller term). In the transition, the $\pi$ bonds become anti-bonding $\pi^*$ under the electric transition moment $\boldsymbol   e\cdot \boldsymbol r$ acting on each bond, causing them to phase (sign) match the corresponding conduction band orbitals and giving rise to a non-vanishing local transition moment.  The local moments are summed to give the total transition moment and are seen to be modulated \textit {at twice the wavevector of the electronic $q$},  along the backbone of the molecule. 
   The modulation is caused by  the oscillation in orbital occupation, with Bloch vector $q$.  The   modulation for this $q$ is seen on the right, shown  with blue arrows. The arrows  point in the same direction but oscillate in strength with a  $k=2q$ periodicity.   This modulation tends to generate a $k=2q$ phonon upon electron-hole pair formation; since $q$ changes with photon h$\nu$ according to the electronic band structure, this is  responsible for Raman  sideband  dispersion. Since  $\cos(q x)^2 = \frac{1}{2} (1 + \cos(2q x))$, there is   a  $k=0$   constant or ``DC''  component.  This sets up creation of a $\Gamma $-point phonon, independent of the excitation frequency or $q$; i.e. the $k=0$ line is always present and has no dispersion.  This story plays out for each of the Raman active modes. }
   \label{fig:polyacetylene3}
\end{figure}

 Figures~\ref{fig:polyacetylene1} and  \ref{fig:polyacetylene3} show a top-down representation of a portion of an infinite length quasi-1D polyacetylene crystal.  
The lowest and highest   extremal $q=0$ (crystal momentum zero)  $\Gamma$ point states, and a representative intermediate state  for the valence band and conduction bands are shown. The valence band electronic states exclusively consist of bonding $\pi$ orbitals (same sign on both carbon atoms) on the double bonded carbons, and the conduction bands   exclusively consist of anti-bonding $\pi$ orbitals (opposite sign) on the double bonded carbons.  The figures and their captions make clear that  $k=0$ and $k=2q$ phonon production is expected, where $q$ is the Bloch wave vector of the conduction band electronic orbital created  by the incoming photon. The sideband intensity will turn out to depend on electron backscattering (see below), but the essence of the sideband dispersion is already clear, as a result of electronic state dispersion resulting in wave vector $q$, and phonon dispersion resulting in an energy shift of the appropriate $j^{th}$ band, $\epsilon_j(k=2 q)$.

\textbf{Transition moments and Herzberg-Teller expansion}
The  valence band orbital  $\psi_{  q}^v({ \boldsymbol \xi} ; \boldsymbol{ r})$ possesses Bloch vector $ q$ (reducing to one dimensional notation for $q$ for the pseudo 1D crystal), phonon coordinates $ { \boldsymbol \xi}$, and electron coordinates $\boldsymbol{ r}$.  The conduction band state $\psi_{ q}^c({ \boldsymbol \xi} ; \boldsymbol{ r})$  has the same Bloch vector (in the case of the $k=0$ band, or a reversed Bloch vector $-  q$ if the Herzberg-Teller term creates a phonon of wavevector $k=2 q$). The transition moment for polarization $\hat  {\boldsymbol  e}_i$ is given by the  integral over electrons
\begin{equation}
\label{mu}
\mu_{qq}^{(vc,  { \boldsymbol  e}_i)}({ \boldsymbol \xi})\equiv \mu_{qq}^{ { \boldsymbol  e}_i}({ \boldsymbol \xi}) =\langle \psi_q^c({ \boldsymbol \xi}; \boldsymbol {r})\vert ( {   \boldsymbol  e}_i\cdot \boldsymbol {r})\vert \psi_q^v({ \boldsymbol \xi}; \boldsymbol {r})\rangle_{\boldsymbol{ r}}.
\end{equation}
The    phonon   coordinates  $ \boldsymbol \xi$ can be used   to form  phonon wave functions, e.g.  $\vert \phi_{j,q}\rangle \propto { \xi}_{j,q} \vert \phi_{0}\rangle$, where  $\vert \phi_{0}\rangle$ is the ground vibrational wave function (zero phonon occupation) of the whole lattice, and ${ \xi}_{j,q}$ is a phonon coordinate with wave vector $q$ in the $j^{th}$ band.
Multiplication by the transition moment $\mu_{qq}^{ {\boldsymbol  e}_i}({ \boldsymbol \xi})$ produces
\begin{equation}
\phi_0(\boldsymbol \xi)\xrightarrow{h\nu} \mu_{qq}^{ {\boldsymbol  e}_i}({ \boldsymbol \xi}))\phi_0(\boldsymbol \xi) =\left (\mu_{qq}^{ {\boldsymbol  e}_i}({ \boldsymbol \xi}_0)+\sum_{j,k}\frac{\partial \mu_{qq}^{ {\boldsymbol  e}_i}({ \boldsymbol \xi})}{\partial \xi_{j,k}}\Bigg |_{\boldsymbol \xi_0} \cdot  \xi_{j,k} +\dots\right )\phi_0(\boldsymbol \xi);
\end{equation}
the RHS is the Herzberg-Teller expansion.   This implies  \textit  {instant phonon creation}, since
\begin{eqnarray}
  \mu_{qq}^{ {\boldsymbol  e}_i}({ \boldsymbol \xi})\phi_0 ( \boldsymbol \xi)  
&=& \mu_{qq}^{ {\boldsymbol  e}_i}({ \boldsymbol \xi_0})\phi_0 ( \boldsymbol \xi) + \sum\limits_{j,k}\left (\frac{\partial \mu_{qq}^{ {\boldsymbol  e}_i}}{\partial  { \xi}_{j,k}} \right )\cdot {\xi}_{j,k}\ \phi_0 ( \boldsymbol \xi) \\ \nonumber &+&\frac{1}{2}\sum\limits_{j,k;j'k'}\left (\frac{\partial^2 \mu_{qq}^{ {\boldsymbol  e}_i}}{\partial {  \xi}_{j,k}\partial {  \xi}_{j',k'} }\right )\cdot { \xi}_{j,k}{  \xi}_{j',k'}\ \phi_0( \boldsymbol \xi) + \cdots  \\ \nonumber
&=& a_0 \ \phi_0 ( \boldsymbol \xi) + \sum\limits_{j,k} b_{j,k} \ \phi_{j,k}( \boldsymbol \xi)  + \cdots  \\
\nonumber
&=& a_0 \ \vert  \boldsymbol 0\rangle   + \sum\limits_{j,k} b_{j,k} \  \vert  k_j\rangle +  \sum\limits_{j,k;j',k'} c_{j,k;j',k'} \  \vert   k_j,k'_{j'}\rangle + \cdots  
\end{eqnarray}
after switching to Dirac notation.
This is a sum over the ground state and excited phonon modes $\vert  k_j\rangle,  \vert   k_j,k'_{j'}\rangle, \cdots$, including possible multiple occupation of the same mode (overtones). However, in a perfect crystal, all of the $ \partial \mu_{qq}^{ {\boldsymbol  e}_i}/\partial  { \xi}_{j,k}$ vanish unless $k=0$, since in carrying out the integral in equation~\ref{mu} the Bloch wave $\exp[i q r]$ is cancelled by $\exp[-i q r]$ due to the complex conjugate in the integral. This leaves any phonon   Bloch oscillation uncompensated, causing the integral to vanish (simply a manifestation  of crystal momentum conservation). Thus the $k=0$ bands and their fixed positions, independent of incident wavelength or $q$,  are explained.

For the dispersive sidebands we need to consider a different transition moment, $\mu_{-q,q}$ or $\mu_{q,-q}$; this  gives a factor $\exp[\pm 2i q r]$ in the   integral, requiring $k=\pm 2 q$ to compensate.  This is  the  genesis of the $k=\pm 2 q$ sidebands, as already implied by the simple tight binding model above. Then   $ \partial \mu_{q,-q}^{ {\boldsymbol  e}_i}/\partial  { \xi}_{j,k=2q}$ does not vanish.  The phonon dispersion curve determines where the sideband peak for  $k=\pm 2 q$ falls in Raman displacement.   However, in a crystal, the electron pseudomomentum is reversed upon promotion to the conduction band and cannot yet recombine with the hole.  It (or the hole) needs to elastically backscatter off defects or an end of the molecule; the $k=2q$ phonon   present   could in principle also backscatter electrons elastically, but experiments    show the $k=2q$ sideband intensity effectively vanishing if artificial sources of backscattering are absent (see below).

With modest concentrations of defects and presence of ends, $k$ and $q$ are     no longer  good quantum numbers, making $k$'s close to $0$ allowed.  Defects play a dual role in the  $k=\pm 2 q$ sideband, also making nearby $k$'s available and  backscattering electrons so they can fill the holes they created. 

\textbf{Herzberg-Teller strength estimates}  The Herzberg-Teller expansion is simply a Taylor expansion of the coordinate dependence that comes whole within Franck-Condon theory;  it is not a perturbation expansion. The  Raman process remains  second order in perturbation theory, including nonperturbative production of phonons.  
 
 There is indeed a strong dependence of the propensity to make a  $\pi$ to $\pi^*$  transition depending on carbon interatomic distance, in a single bond. However does this survive the transformation to a finite derivative with respect to phonon coordinates?  Detailed arguments are given in the supplementary materials show this is the case, but there is a very quick and convincing shortcut to the conclusion: if the phonon coordinate derivatives vanished, the off resonance Raman scattering would too: the Placzek polarizability derivative\cite{soo} with phonon  coordinate would vanish for all phonons. This is obviously not the case (see figure~\ref{fig:poly}).

If a phonon   of wave vector $k=2q,$ is created instantaneously upon excitation, the energy devoted to the electronic transition is adjusted by the phonon energy, according to the total energy in the photon: $E_{\textrm{phonon}}+ E_{\textrm{electron transition}} = h \nu$.  It retains the matching $\vert q\vert$ if the electron or hole do \textit{not} experience electron-phonon scattering.  The phonon's energy dispersion is thus written into the Raman sideband dispersion.  

 The $k=2q$ phonon sidebands diminish in strength (and move toward the $k=0$ line) with redder incoming light and are missing altogether below resonance.  If  the phonons are reliably produced as a by-product of the transition moment, why do  corresponding Raman sidebands diminish in intensity this way?  Off resonance, there is no time to backscatter  electrons, so the  electrons, requiring backscattering to emit, remain very unlikely to find their way back to the hole they left behind.  There is no such problem for the $k=0$ band, since the electron did not change crystal momentum in the first place.  These facts   contribute to the large change in the ratio of $k=0$ and $k=2q$ Raman band intensities with incident frequency. 
In summary, the phonons are reliably created, but the fraction that gives rise to Raman shifted emission depends on backscattering conditions.

 \begin{figure}[htbp] 
    \centering
    \includegraphics[width=6in]{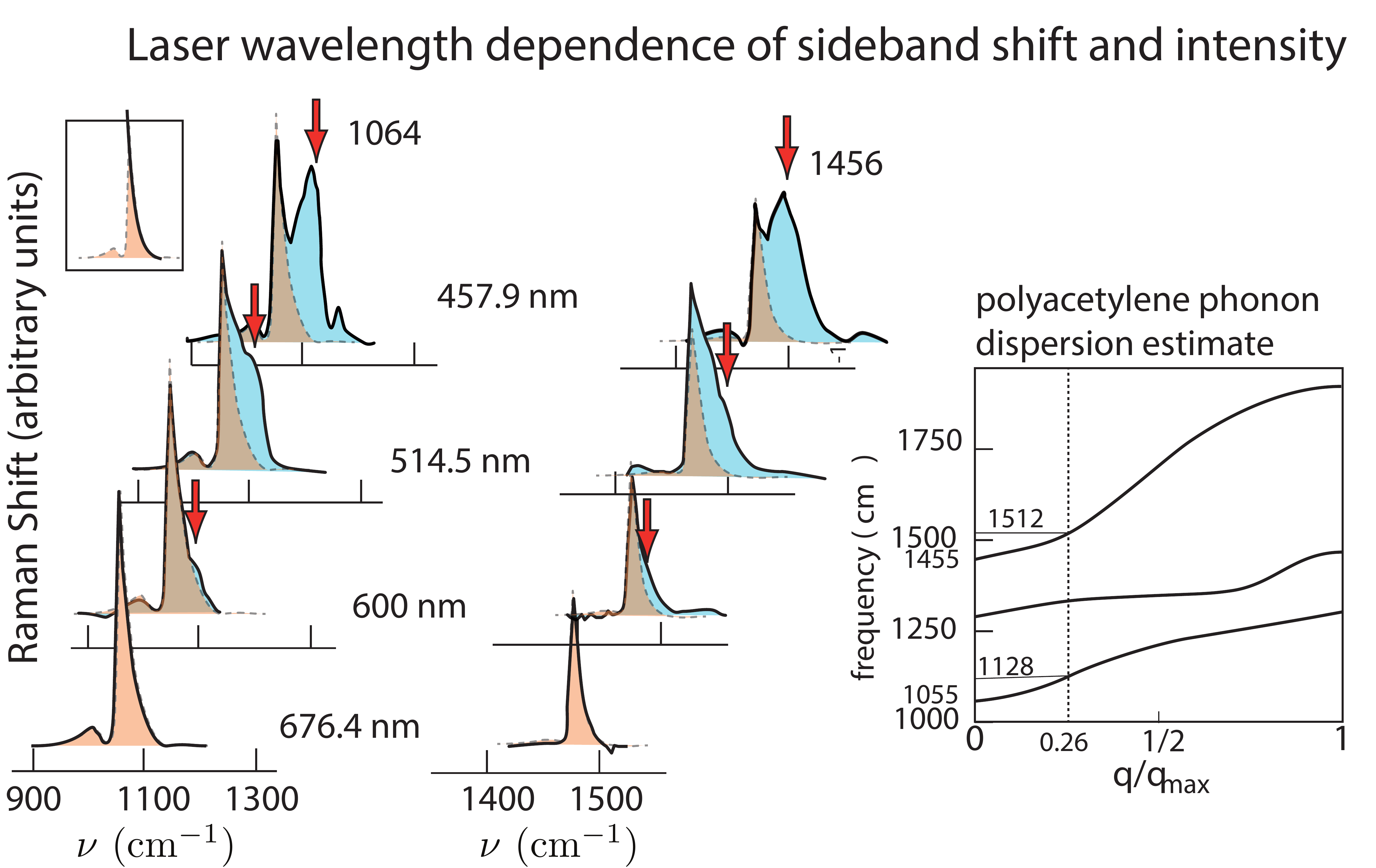} 
    \caption{The dispersion  and growth 	 of the Raman sidebands  of  the 1064 cm$^{-1}$ and 1456 cm$^{-1}$ sidebands of \textit{trans}-polyacetylene taken by Mulazzi \textit{et. al.}  at the laser frequencies shown, redrawn from \cite{junk1}.   Our prediction of sideband position,  using the phonon bands in the inset, right, is shown by the red arrows. (We have not found a well established, ``most reliable" phonon dispersion for polyacetylene. One can say  we have here established the phonon dispersion in the early part of the bands for the first time, through   interpretation of the experiments. 
    The overall line shape is the sum of the  constant $k=0$ band and the moving and  growing $k=2q$ sideband.  The latter  is created  by defects and ends, i.e. causing the non-periodic part of the electronic and vibrational states.  The inset (upper left) shows the $k=0$ band fit to an exponential fall-off on the right for the 1064 cm$^{-1}$ band. Inset,  right: Polyacetylene phonon dispersion curves according to Jumaeu \textit{et.al.}\cite{jumeau}. 
    }
    \label{fig:multifit}
 \end{figure}

\textbf{Defects and finite sized molecules} 
What makes the  features of the polyacetylene Raman band shapes peculiar, beyond the dispersion of $k=2q$ Raman bands, is the strongly variable strength of the (broadened)  sidebands depending on conditions, and the  width and shape of the band. These are not Gaussian or Lorenzian lineshapes living under a sum rule!
The band shapes and their evolution with incident wavelength can be explained in terms of the different responses of the $k=0$ peak and the $k=2q$ peak to the effects of backscattering.  

\begin{figure}[htbp] 
   \centering
   \includegraphics[width=6 in]{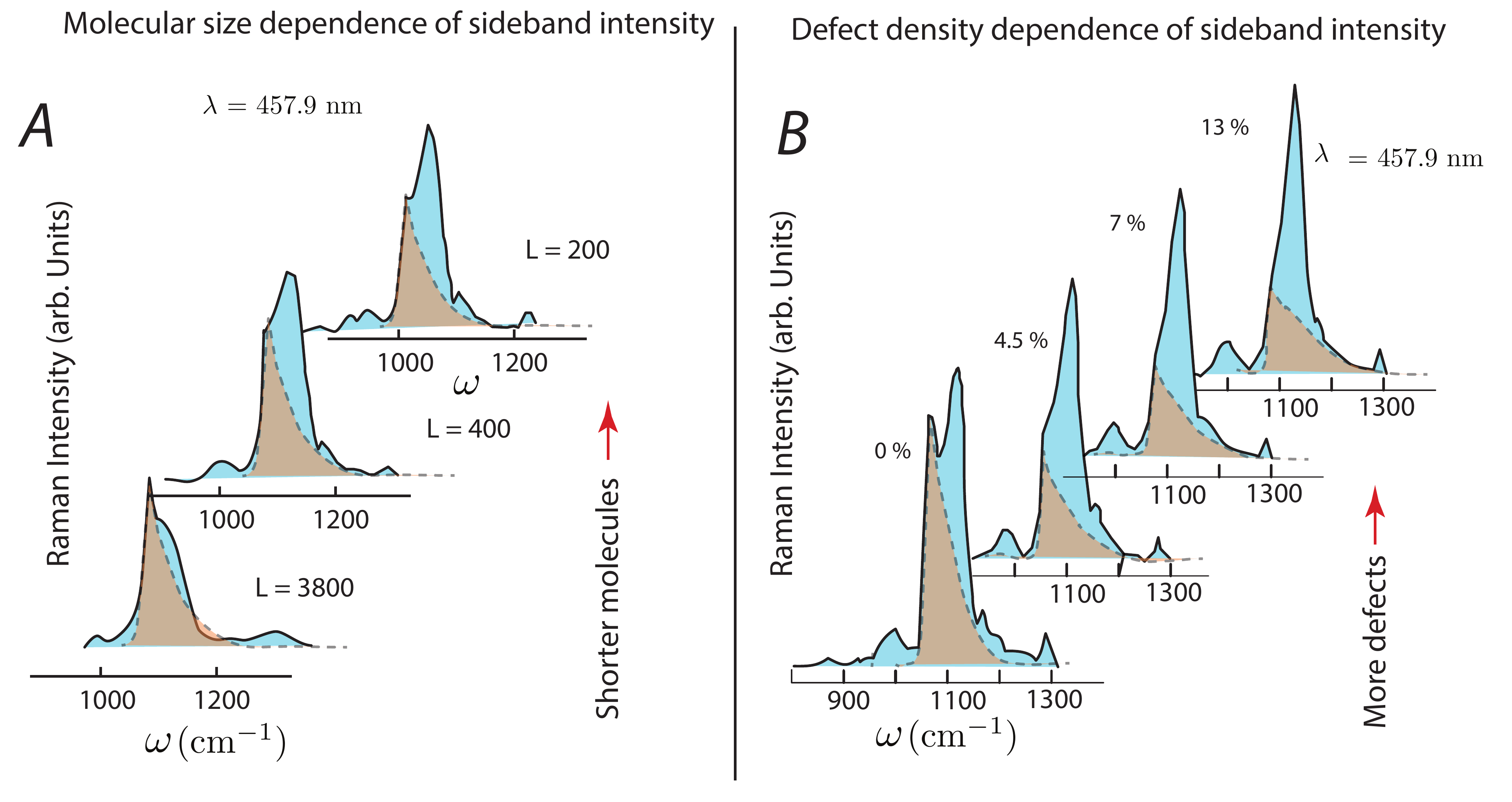} 
   \caption{A. As the polyacetylene length becomes shorter in a monodisperse sample, the ends of the molecule become accessible to a higher fraction of the electrons, increasing the backscattering efficiency. Note the important details: 1) the sideband increases in intensity but does not broaden in FWHM after 4.5\% defects. The sideband width and line shape is not lifetime determined.  2) The $k=0$ band is of course added to the sideband to give the total intensity, making a break in slope of the total to the right of the $k=0$ band  peak in both A and B panels (and verified in the numerical calculations, figure!\ref{fig:simuspec}.   The prediction  that the $k=2d$ intensity \textit{falls} like the inverse length of the molecule (assuming no defect scattering) is supported by the evolution of the ratio of the   $(k=0)$ to $(k=2q)$ band intensity ratio, using the same $k=0$ band shape (dashed line) as in figure~\ref{fig:multifit}.  This ratio  is 1:7.5.  B. As the density of defects increases, the $k=2q$ dispersive sideband accounts for an increasing fraction of the Raman scattering in the 1054 cm$^{-1}$ band. Other  sideband features in different bands show the same behavior. The figure was redrawn and the dashed $k=0$ band contribution added, starting from Sch\"afer-Seibert \textit{et. al.}\cite{defects}.  }
   \label{fig:defect}
\end{figure}

  The $k=0$ $\Gamma$ point bands are always present for Raman allowed transitions, induced by the constant component of transition moment; these don't require backscattering to in order to be produced.  The exponential tail to the right of the sharp $k=0$ feature at 1164 cm$^{-1}$ is found not to depend on backscattering strength (pink and tan shaded regions, figures~\ref{fig:multifit},\ref{fig:defect}).  If defects and ends are present, $k$ is no longer a good quantum number. The degeneracy of left and right traveling plane waves  is broken, both electronically and vibrationally; the phonons associated with the $k=0$ line carry no pseudomomentum even as they carry energy above that of the $k=0$ line.  The energies do not lie below the $k=0$ line because  no phonon states exist there. Even with defects present, it is difficult to generate vibrations of lower frequency than the $\Gamma$ point of each band, since confining the vibrations tends to produce  higher, not lower frequencies if the band dispersion slope is positive, as it is here. Thus the abrupt fall-off to the left of the $\Gamma$ point line. 


It is remarkable that  long ago three critical experimental tests  were performed that  support the transition moment/backscattering model given here: varying the incident wavelength, varying the length of the molecule in a controlled way \cite{lengths}, and varying the defect density in a controlled way.

In reference \cite{lengths}, three samples of nearly mono disperse polyacetylene   with lengths of about 200, 400, and   3800  unit cells were produced and their Raman spectra taken. Many of the  earlier explanations for the line shape thus evaporated. 

 A shorter  polyacetylene molecule has ends   available  to backscatter to  a larger fraction of electrons.    The prediction is that the $k=2d$ intensity falls like the inverse length of the molecule assuming no defect scattering. The ratio of the larger to smallest $(k=0)/(k=2q)$ ratio in figure~\ref{fig:defect}, left, using the same $k=0$ band shape (dashed line) as in figure~\ref{fig:multifit} is 1:7.5.  Assuming the (a crude guess) 100 unit cell proximity rule to backscatter, the ratio should  have been roughly 1:20.  The reduced effect of the end proximity reflects the error in the coherence length guess, or residual defect backscattering effects, or, more likely, both.  In any case, accessible ends for backscattering dramatically enhance the $k=2q$ band, according to both the model and the experiment (see figure~\ref{fig:defect}.)

Another key measurement involved controlled oxidation of the polyacetylene, resulting in a knowable \textit{additional} defect density (over the nascent density)  of 0\%, 4.5\%, 7\%, or 13\%. figure~\ref{fig:defect}.   The ratio of the larger to smallest $(k=0)/(k=2q)$ ratio is about 1:6, meaning six times as many electrons relax by backscattering, emitting and filling their holes with the highest defect density compared to nascent density plus end effects.

\begin{figure}[htbp] 
   \centering
   \includegraphics[width=6.5in]{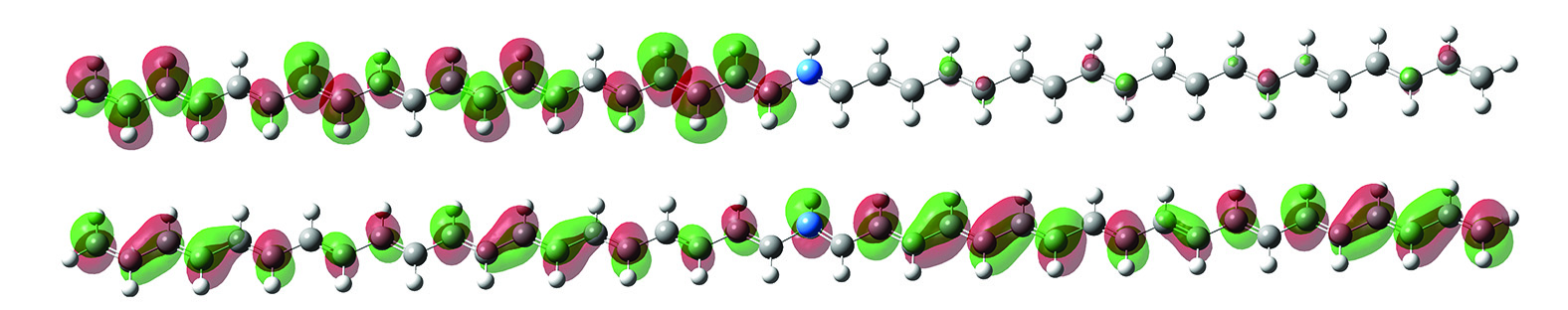} 
   \caption{At the bottom of the conductance band, at long Bloch wavelength, the electronic wave function (bottom) in this Gaussian  09 DFT calculation at the 3-21G +  level substantially ignores the (blue) Si atom defect, but at shorter Bloch wavelength (top) the backscattering is more severe,   localizing the eigenstate to mostly one side.  }
   \label{fig:defectrend}
\end{figure}

Finally we discuss the trends with laser frequency, as seen in figure~\ref{fig:multifit}. Preliminary  calculations using 20 unit cell (40 carbon atom) polyacetylene molecules with defects caused by Si replacing C, or oxidation giving a carbonyl in place of a normal chain carbon both show a general trend toward increased backscattering with increased $k$ (figure~\ref{fig:defectrend}); this trend would support the growth of the sideband area with shorter laser wavelengths. However a more significant trend may be the growth in the number of possible sideband transitions as dispersion opens a larger gap (on the order of 50 cm$^{-1}$) between the $k=0$ and $k=2 q$ peaks.  More states become available to be populated with phonons.  Our calculations show that Rayleigh scattering is still the dominant process, so there is much leeway for phonon production to become a large fraction of the results of photoabsorption. 

\begin{figure}[htbp] 
   \centering
   \includegraphics[width=6.5in]{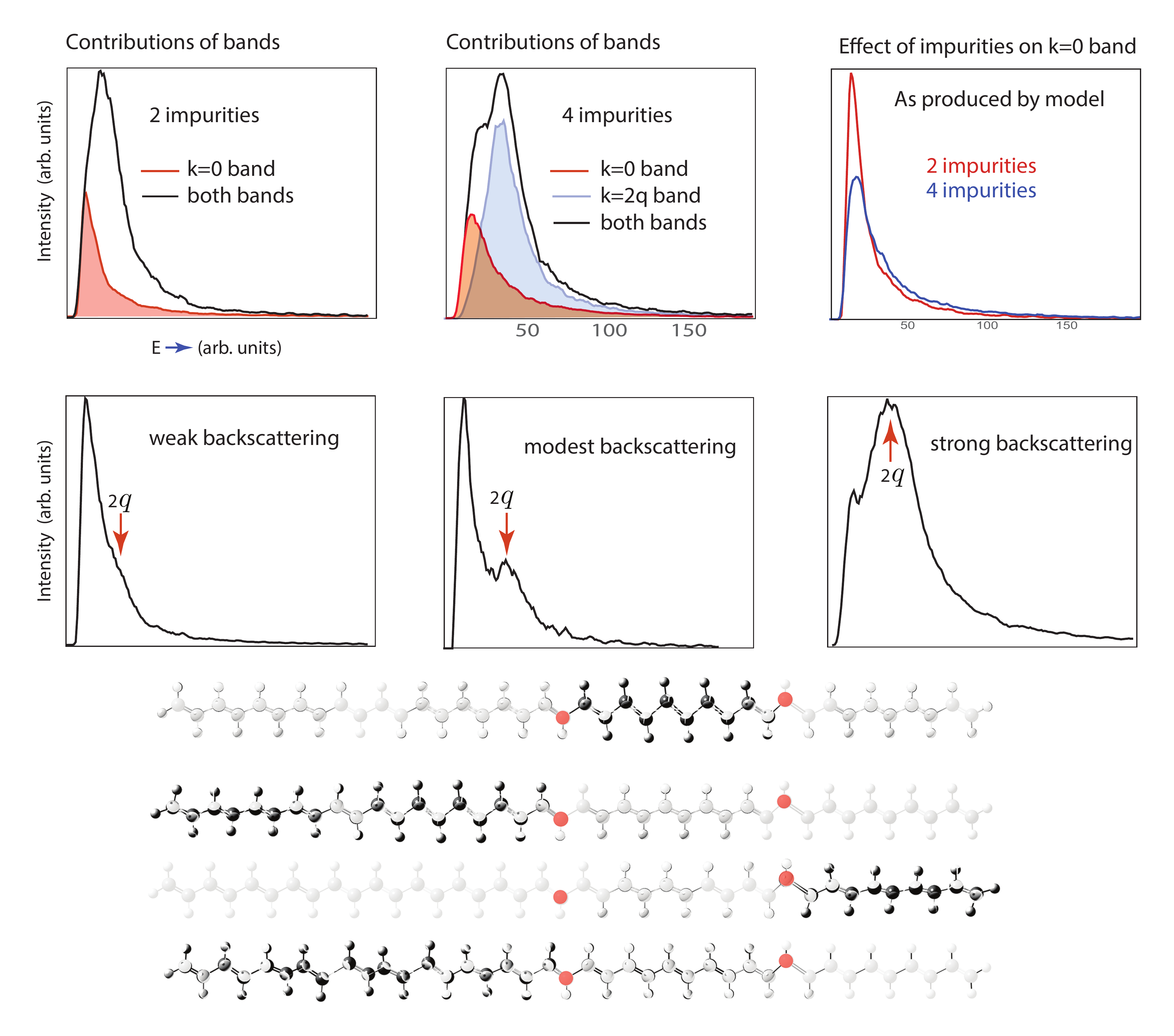} 
   \caption{Numerical results from a simple phonon tight binding simulation of the Raman theory presented in this paper.  The calculations involve a 450 unit cell long   sample with alternating ``bonds'' and two or four randomly placed defect impurities. About 3000 placements of the random impurities were averaged here. The  $k=0$ and $k=2q$ components can be computed separately (see their contribution to the total in the top row), and their ratio was   varied by hand, since we could not do an accurate simulation of the electron localization and backscattering in such a simple model. The lines shapes nonetheless emerge naturally from the simulation and are due to the partial vibrational confinement between ends and defects, seen at the bottom in the simulation and in a Gaussian 09 density functional calculation with four Si atoms in place of carbon. Four low lying vibrational modes are seen, with darkness of the atoms representing vibration amplitude.  The vibrational modes have a near linear  dispersion in this low $k$ regime and are driven by the constant and sinusoidal terms in the transition moment, as in the tight binding model presented above in figures~\ref{fig:polyacetylene1} and \ref{fig:polyacetylene3}.}
   \label{fig:simuspec}
\end{figure}


As a check on the mechanisms for lineshape evolution, we constructed a tight binding model with the molecular backbone represented by alternating   bonds, of length 450 unit cells, including 2 or 4 randomly placed 10\% mass defects.  The spectra were calculated for each case by assuming   regular, unperturbed constant and sinusoidal driving terms coming from transition moment coordinate dependence of the electronic transitions (we find the electronic states to be much less sensitive to impurities than the vibrations (phonons), which readily semi-localize in zones between impurities, according to our Gaussian 09 DFT calculations).  The spectrum was computed by calculating the excitability of each phonon mode of the chain at the given driving ``qÕ' (with impurities in place), and adding its contribution to  the Raman shift spectrum at the mode's frequency.  The sideband backscattering intensity was adjusted by hand and various limits of high and low impurity, long and short molecules, etc. investigated this way. The results for various combinations are shown in figure~\ref{fig:simuspec}, which should be compared with figures~\ref{fig:multifit},\ref{fig:defect}, and \ref{fig:poly}.  It takes  about 3000 random realizations of the impurity positions before the average (shown) settles down.

\textbf{Implications and conclusion}
The spectrum of polyacetylene has been explained, in terms of Franck-Condon theory without the Condon approximation.  The key  is the transition moment and its coordinate dependence, leading to  immediate phonon production. In an infinite crystal with delocalized orbitals there is also the absence of new forces in the excited states.  Raman sidebands, their dispersion and lineshapes are now understood, requiring backscattering to exist. The theory applied here  will be widely applicable to other systems, including graphene.  There is a major shift of emphasis from phonons produced after the fact of photoabsorption, to phonons produced instantaneously upon photoabsorption. 

For the future, it is important to perform high level electronic structure to map out the transition moment as a function of atomic or phonon displacement. It is important, if difficult perhaps, to experimentally check for the immediate presence of phonons after photoabsorption.  This cannot be done by looking for immediate Raman emission in the sidebands, since backscattering must occur first, but the $k=0$ bands should not suffer this, implying an evolution of the sidebands with time in a pulsed experiment. 

\textbf{Acknowledgements} The authors acknowledge support from the NSF Center for Integrated Quantum Materials(CIQM) through grant NSF-DMR-1231319. We also thank Prof. Philip Kim for helpful conversations.



\end{document}